\documentclass[aps,prb,twocolumn,showpacs,preprintnumbers,amsmath,amssymb,superscriptaddress]{revtex4}
\usepackage{mathptm}  
\usepackage{dcolumn}                    
\usepackage{bm}                        
\usepackage{graphicx}
\usepackage{times}
\usepackage{epstopdf}
\begin{document}

\newcommand{\Imag}{{\Im\mathrm{m}}}   
\newcommand{\Real}{{\mathrm{Re}}}   
\newcommand{\im}{\mathrm{i}}        
\newcommand{\talpha}{\tilde{\alpha}}
\newcommand{\ve}[1]{{\mathbf{#1}}}

\newcommand{\x}{\lambda}  
\newcommand{\y}{\rho}     
\newcommand{\T}{\mathrm{T}}   
\newcommand{\Pv}{\mathcal{P}} 
\newcommand{\vk}{\ve{k}} 
\newcommand{\vp}{\ve{p}} 

\newcommand{\N}{\underline{\mathcal{N}}} 
\newcommand{\Nt}{\underline{\tilde{\mathcal{N}}}} 
\newcommand{\g}{\underline{\gamma}} 
\newcommand{\gt}{\underline{\tilde{\gamma}}} 

\newcommand{\vecr}{\ve{r}} 
\newcommand{\vq}{\ve{q}} 
\newcommand{\ca}[2][]{c_{#2}^{\vphantom{\dagger}#1}} 
\newcommand{\cc}[2][]{c_{#2}^{{\dagger}#1}}          
\newcommand{\da}[2][]{d_{#2}^{\vphantom{\dagger}#1}} 
\newcommand{\dc}[2][]{d_{#2}^{{\dagger}#1}}          
\newcommand{\ga}[2][]{\gamma_{#2}^{\vphantom{\dagger}#1}} 
\newcommand{\gc}[2][]{\gamma_{#2}^{{\dagger}#1}}          
\newcommand{\ea}[2][]{\eta_{#2}^{\vphantom{\dagger}#1}} 
\newcommand{\ec}[2][]{\eta_{#2}^{{\dagger}#1}}          
\newcommand{\su}{\uparrow}    
\newcommand{\sd}{\downarrow}  
\newcommand{\Tkp}[1]{T_{\vk\vp#1}}  
\newcommand{\muone}{\mu^{(1)}}      
\newcommand{\mutwo}{\mu^{(2)}}      
\newcommand{\epsk}{\varepsilon_\vk}
\newcommand{\epsp}{\varepsilon_\vp}
\newcommand{\e}[1]{\mathrm{e}^{#1}}
\newcommand{\dif}{\mathrm{d}} 
\newcommand{\diff}[2]{\frac{\dif #1}{\dif #2}}
\newcommand{\pdiff}[2]{\frac{\partial #1}{\partial #2}}
\newcommand{\mean}[1]{\langle#1\rangle}
\newcommand{\abs}[1]{|#1|}
\newcommand{\abss}[1]{|#1|^2}
\newcommand{\Sk}[1][\vk]{\ve{S}_{#1}}
\newcommand{\pauli}[1][\alpha\beta]{\boldsymbol{\sigma}_{#1}^{\vphantom{\dagger}}}

\newcommand{\eq}{Eq.}
\newcommand{\eqs}{Eqs.}
\newcommand{\cf}{\textit{cf. }}
\newcommand{\ie}{\textit{i.e. }}
\newcommand{\eg}{\textit{e.g. }}
\newcommand{\etal}{\emph{et al.}}
\def\i{\mathrm{i}}

\title{Interplay between Superconductivity and Ferromagnetism on a Topological Insulator}

\author{Jacob Linder}
\affiliation{Department of Physics, Norwegian University of
Science and Technology, N-7491 Trondheim, Norway}

\author{Yukio Tanaka}
\affiliation{Department of Applied Physics, Nagoya University, Nagoya, 464-8603, Japan}

\author{Takehito Yokoyama}
\affiliation{Department of Physics, Tokyo Institute of Technology, 2-12-1 Ookayama, 
Meguro-ku, Tokyo 152-8551, Japan
}

\author{Asle Sudb{\o}}
\affiliation{Department of Physics, Norwegian University of
Science and Technology, N-7491 Trondheim, Norway}

\author{Naoto Nagaosa}
\affiliation{Department of Applied Physics, University of Tokyo, Tokyo 113-8656, Japan}
\affiliation{Cross Correlated Materials Research Group (CMRG), ASI, RIKEN, WAKO 351-0198, Japan}

\date{\today}

\begin{abstract}

We study theoretically proximity-induced superconductivity and ferromagnetism on the surface of a topological insulator. In particular, we investigate how the Andreev-bound states are influenced by the interplay between these phenomena, taking also into account the possibility of unconventional pairing. We find a qualitative difference in the excitation spectrum when comparing spin-singlet and spin-triplet pairing, leading to non-gapped excitations in the latter case. The formation of surface-states and their dependence on the magnetization orientation is investigated, and it is found that these states are Majorana fermions in the $d_{xy}$-wave case in stark contrast to the topologically trivial high-$T_c$ cuprates. The signature of such states in the conductance spectra is studied, and we also compute the supercurrent which flows on the surface of the topological insulator when a Josephson junction is deposited on top of it. It is found that the current exhibits an anomalous current-phase relation when the region separating the superconducting banks is ferromagnetic, and we also show that in contrast to the metallic case the exchange field in such a scenario does not induce 0-$\pi$ oscillations in the critical current. Similarly to the high-$T_c$ cuprates, the presence of zero-energy surface states on the topological surface leads to a strong low-temperature enhancement of the critical current.  

\end{abstract}
\maketitle

\section{Introduction}

Topological insulators \cite{konig_jpsj_08, kane_prl_05, bernevig_06, 3d} represent a state of matter which is distinct from all other condensed matter systems. Their hallmark is the formation of topologically protected, conducting edge-states (in 2D) and surface-states (in 3D), whereas the bulk retains an insulating behavior (see \eg Ref.~\onlinecite{review_TI} for a nice introduction). These surface-states are characterized by a topological $Z_2$ symmetry and are robust against disorder and perturbations that respect time-reversal symmetry. The key to this robustness is the fact that the Brillouin zone of topological insulators feature an \textit{odd} number of Dirac cones (in contrast to the \textit{even} number of such cones in graphene), which ensures that backscattering paths due to \eg impurities always interfere destructively. 

After their prediction \cite{bernevig_06}, topological insulators have been experimentally observed in HgTe/CdTe quantum wells \cite{konig_jpsj_08} and in Bi$_2$Se$_3$/Bi$_2$Te$_3$ crystals \cite{hsieh}. Recently, several works have investigated proximity-induced superconducting and ferromagnetic order on the surface of a topological insulator \cite{fu_prl_09, akhmerov_prl_09, tanaka_prl_09, law_arxiv_09, santos_arxiv_09, volovik_jetp_09}. It has been found that the synthesis of the abovementioned orders in the environment offered by the topological insulator yield a number of interesting possibilities. On the one hand, it has been shown that such hybrid structures can host so-called Majorana fermions \cite{majorana}. This class of excitations, in contrast to their Dirac equivalent, are their own antiparticles and satisfy non-Abelian statistics \cite{fu_prl_08}. The latter aspect, in similarity to the fractional quantum Hall effect, has prompted suggestions of using such excitations in topological quantum computation owing to their robustness towards decoherence effects. It should be noted that it recently has been proposed that Majorana excitations may be generated in semiconductor$\mid$superconductor heterostructures in the presence of a magnetic field \cite{sau_prl_10, alicea_prb_10}. On the other hand, the study of topological insulators also attracts interest due to the possibility of unveiling novel transport phenomena with respect to spin- and charge-transport \cite{yokoyama, mondal_prl_10}.
 
In Ref. \cite{linder_prl_10}, it was recently shown how an interplay between unconventional superconductivity and ferromagnetism on the surface of a topological insulator would give rise to a number of effects with no counterpart in conventional metallic systems. For instance, the proximity-effect from a spin-triplet superconductor would give rise to gapless excitations in the topological insulator since the gap simply renormalized the chemical potential. Moreover, it was demonstrated how the zero-energy states \cite{hu_prl_94, tanaka_prl_95} formed due to a $d_{xy}$-wave order parameter were Majorana fermions in contrast to \eg the topologically trivial high-$T_c$ cuprates, and that the dispersion of these states would be highly sensitive to the orientation of a magnetic field. All of these findings demonstrate that qualitatively new effects may be expected when superconducting and ferromagnetic order conspire in the environment of a topological insulator.

\begin{figure}[t!]
\centering
\resizebox{0.48\textwidth}{!}{
\includegraphics{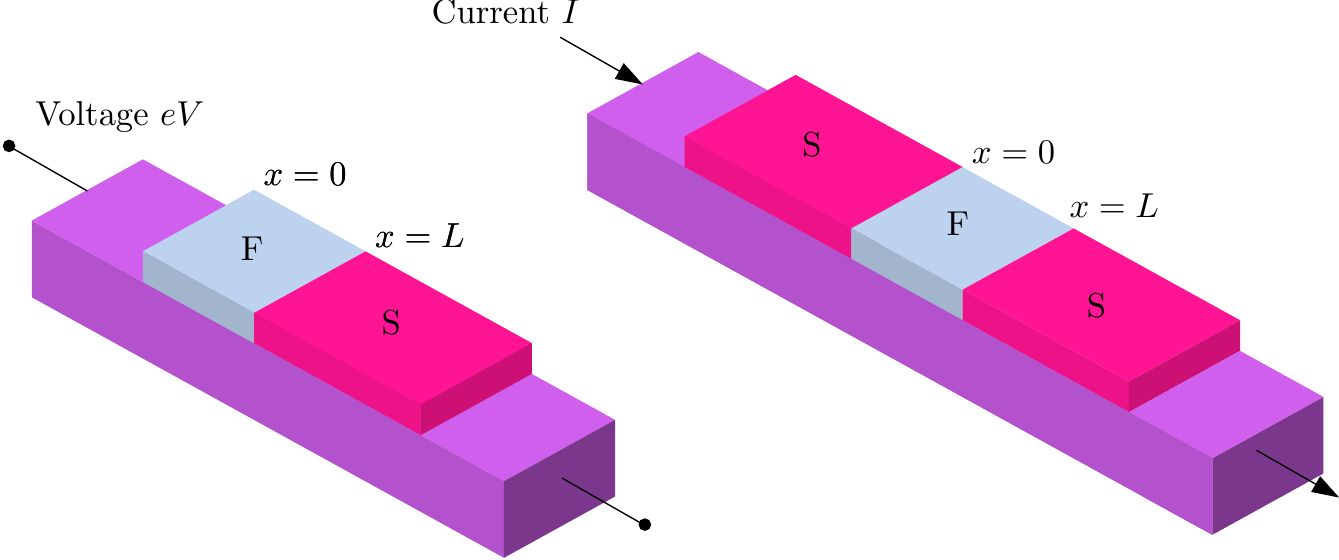}}
\caption{(Color online) We consider a topological insulator where superconductivity and/or magnetic correlations are induced on the surface via the proximity effect to host materials with the desired properties. 
The possibility of unconventional pairing such as $d$-wave is taken into account, and the degeneracy of the band-structure is lifted in the presence of an exchange field induced from the ferromagnetic insulator.
}
\label{fig:model} 
\end{figure}

Motivated by this, in this work we present a comprehensive treatment of hybrid superconductor$\mid$ferromagnet structures deposited on top of a topological insulator. In particular, we focus on their transport properties and how the Andreev reflection process is altered compared to in conventional metallic systems. We study both point-contact spectroscopy and Josephson junction geometries which are directly experimentally relevant, and allow for the possibility of unconventional superconducting pairing such as $p$-wave or $d$-wave. Our proposed model is shown in Fig. \ref{fig:model}. A voltage or current bias may be applied to a topological insulator where the surface is coated with a ferromagnetic insulator and a superconductor. In this way, one may access both conductance spectra and the supercurrent characteristics in order to probe how these are influenced by the environment of the topological insulator.

The present authors have already published a Letter \cite{linder_prl_10} reporting some of the results in the present paper. In addition to the more detailed and comprehensive explanation of the derivation and results, there are new results presented in Sec. \ref{sec:mix}, \ref{sec:jos}, and in the Appendix of this work.

\section{Theory}

We will employ a Bogolioubov-de Gennes approach to obtain the bound-states and transport properties of the system under consideration. Using a Nambu basis 
\begin{align}
\Psi=(\psi_\uparrow, \psi_\downarrow, \psi_\uparrow^\dag, \psi_\downarrow^\dag),
\end{align}
the Hamiltonian for the surface of a topological insulator under the influence of either a superconducting or magnetic proximity effect reads:
\begin{align}\label{eq:H}
\hat{H} &= \begin{pmatrix}
\underline{H_0}(\vk) + \underline{M} & \underline{\Delta}(\vk)\\
-\underline{\Delta}^*(-\vk) & -\underline{H_0}^*(-\vk) - \underline{M}^*\\
\end{pmatrix},
\end{align}
where we have defined 
\begin{align}
\underline{H_0}(\vk) = v_F(\underline{\sigma_x}k_x +\underline{\sigma_y}k_y) - \mu
\end{align}
and $\underline{\ldots}$ denotes a $2\times2$ matrix. The gap matrix $\underline{\Delta}(\vk)$ depends on both the orbital- and spin-symmetry of the Cooper pair, whereas the ferromagnetic contribution reads $\underline{M} = \mathbf{m}\cdot\mathbf{\sigma}$ with an exchange field $\mathbf{m} = (m_x,m_y,m_z)$. In the superconducting region, we set $\underline{M} = 0$ while in the ferromagnetic region we set $\underline{\Delta}(\vk) = 0$. In the following, we set $v_F=1$. It should be noted that it is also possible to use a Rashba-like term in the normal-state Hamiltonian $\underline{H_0}$, i.e. $(\underline{\sigma_x}k_y -\underline{\sigma_y}k_x)$ instead of $(\underline{\sigma_x}k_x +\underline{\sigma_y}k_y)$. This gives rise to a different spin-momentum locking on the Fermi surface, which was observed experimentally in Ref.~\onlinecite{hsieh_spin}. Nevertheless, all our conclusions below pertaining to the excitation spectrum and the qualitative behavior of the spin-singlet versus spin-triplet superconducting scenarios are independent of whether one uses the Rashba- or Dirac-like model for $\underline{H_0}$.

It is instructive to consider how the band-structure on the surface of the topological insulator depends on the doping level and also the presence of a magnetic field. In the absence of proximity-induced ferromagnetism, the energy dispersion reads:
\begin{align}
\varepsilon = \pm |\vk| - \mu,
\end{align}
which is equivalent to the massless, relativistic Dirac fermions in graphene. However, we underline once again that the Brillouin zone of graphene contains an even number of such Dirac cones whereas that number is odd in a topological insulator. We also note in passing that depending on the magnitude of the chemical potential $\mu$, it is possible to generate so-called specular Andreev reflection \cite{beenakker_prl_06} on the surface of a topological insulator between a region with and without a superconducting gap $\Delta_0$ since the reflected hole may have parallel group velocity and momentum when $\mu \leq \Delta_0$. 

\begin{figure}[t!]
\centering
\resizebox{0.48\textwidth}{!}{
\includegraphics{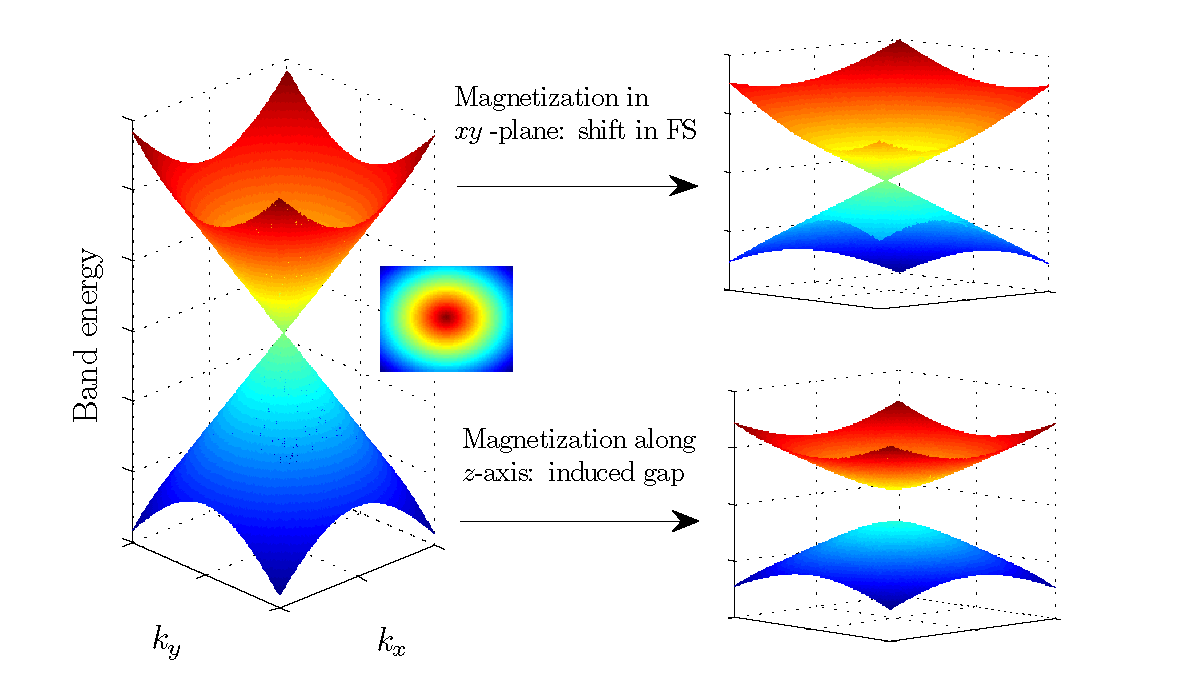}}
\caption{(Color online) The band-structure on the surface of the topological insulator. The helical surface states reside on a Dirac cone, which is influenced in a qualitatively different manner for various magnetization orientations. 
}
\label{fig:properties} 
\end{figure}

In the presence of an exchange splitting induced by a magnetization, the band dispersion becomes:
\begin{align}
\varepsilon = \pm \sqrt{(k_x+m_x)^2 + (k_y+m_y)^2 + m_z^2} - \mu,
\end{align}
It is seen that the exchange field enters in the same way as a vector potential would in a conventional metal. While the transverse components $\{m_x,m_y\}$ of the magnetization shift the position of the Fermi surface, the $z$-component of the field has a qualitatively different effect - it induces a gap in the spectrum. This is seen by considering the $\Gamma$ point $\vk = 0$, where the two bands become separated by an energy gap $2m_z$. The influence of the magnetization on the band-structure is shown schematically in Fig. \ref{fig:properties}. As we shall see later, the direction of the magnetization strongly influences the transport properties of our system. 

We close this section by briefly outlining the relation between the Hamiltonian in Eq. (\ref{eq:H}) and the Hamiltonian of the bulk topological insulator. The Hamiltonian for a 3D topological insulator such as Bi$_2$Se$_3$/Bi$_2$Te$_3$ can be written down by symmetry arguments as shown in Ref. \cite{zhang_nphys_09}. To account for the surface-states, the eigenstates $\Psi$ are obtained by diagonalizing the full Hamiltonian and then demanding that $\Psi=0$ at the boundaries. In this way, one obtains a set of surface-states with a linear energy-momentum dispersion which cross at the $\Gamma$-point. These surface-states can be influenced by finite-size effects \cite{linder_prb_09, lu_prb_10, liu_prb_10} if the topological insulator has a sufficiently small width, typically a few nm in Bi$_2$Se$_3$. In our work, we assume that this width is sufficiently large to rule out any finite-size effects such that the surface-states simply can be described by a Dirac cone, which leads to Eq. (\ref{eq:H}).

\section{Results \& Discussion}

\subsection{Conductance spectroscopy}

In this section, we will focus on the transport properties of hybrid structures deposited on top of a topological insulator in order to investigate how the interplay between ferromagnetism and superconductivity in this context can be probed by conductance spectroscopy. To this end, we will employ a scattering-matrix formulation along the lines of Blonder-Tinkham-Klapwijk \cite{btk} theory, including also the role of unconventional superconducting pairing. To accommodate superconductivity by means of the proximity-effect experimentally, it is necessary to realize the condition $\mu \gg \Delta_0$ to have a sufficiently large density of states. Throughout this paper, we will therefore be concerned with precisely this situation and set $\mu \gg \Delta_0$ everywhere on the surface of the topological insulator except in the ferromagnetic region where we consider $\mu=0$. The reason for this is twofold. Firstly, such a large Fermi-vector mismatch effectively accounts for a barrier between the non-superconducting and superconducting region of the surface of the topological insulator, which is expected to be present experimentally. Secondly, it is necessary to use an insulating material between the voltage source and the superconducting region in order to ensure that transport occurs exclusively along the surface of the topological insulator (and not through the bulk of the host material inducing ferromagnetism). In this way, a ferromagnetic insulator deposited on top of the topological insulator would be suitable experimentally, as depicted in Fig. \ref{fig:model}. We now proceed to discuss the wavefunctions in each of the regions of the surface of the topological insulator.

In the normal (N) region of the surface, we have $\underline{M}=\underline{\Delta}=0$, and an incoming right-moving electron with energy $\varepsilon$ may then suffer two possible fates upon scattering: \textit{i)} reflection as an electron or \textit{ii)} Andreev-reflection as a hole. The total wavefunction may then be written as:
\begin{align}
\psi_N = \e{\i k_y y} \Big([&1,\e{\i\theta},0,0]\e{\i k_x x} + r_e[1,-\e{-\i\theta},0,0]\e{-\i k_x x} \notag\\
&+ r_h[0,0,1,-\e{-\i\theta}] \e{\i k_x x}\Big),
\end{align}
where $\theta$ denotes the angle of incidence while $r_e$ and $r_h$ are the normal and Andreev scattering coefficients, respectively. Here, $k_x = \mu_N\cos\theta$ is the $x$-component of the momentum which is non-conserved due to the broken translational symmetry, whereas $k_y = \mu_N\sin\theta$ is conserved.

\begin{figure}[t!]
\centering
\resizebox{0.3\textwidth}{!}{
\includegraphics{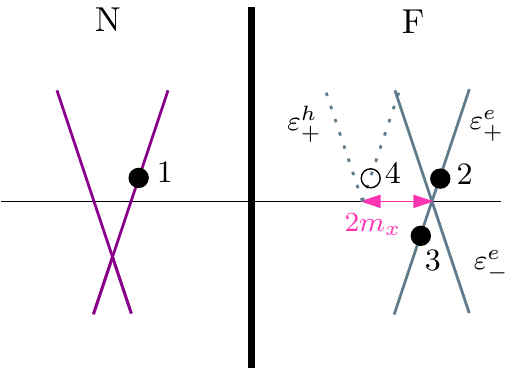}}
\caption{(Color online) Illustration of the scattering processes and the bands partaking in these at the N$\mid$F interface. An incoming electron (1) with $\varepsilon>0$ from the N side can be transmitted (2) to the F region in the band $\varepsilon_+^e$. When this electron hits the S interface (not shown), it has a finite probability of being Andreev-reflected. In this process, an electron is taken (3) from the band $\varepsilon_-^e$ and transmitted to the superconductor, leaving a hole (4) behind. This hole excitation has a band dispersion which is displaced in momentum space with a vector $(2m_x,2m_y)$ compared to the electron-band. Only the $x$-direction is shown above for clarity and we have set $m_z=0$. For non-zero $m_z$, an energy gap $2m_z$ opens between the Dirac-cones in the F region.
}
\label{fig:scattering_NF} 
\end{figure}

In the ferromagnetic insulator (F) region of the surface, we have $\underline{\Delta}=0$ and $\underline{M}\neq0$. We set the chemical potential to zero in this region for reasons outlined above. In that case, the wavefunction in the F region is in general a superposition of right- and left-moving electrons, due the transmission from the N region and reflection at the S interface, in addition to right- and left-moving holes, due to Andreev-reflection at the S interface and normal reflection at the N interface. Thus, we obtain:
\begin{align}\label{eq:F}
\psi_F = \e{\i k_y y} \Big(&a_1[\i\alpha_+, 1, 0, 0]\e{-(\kappa_+ + \i m_x)x} \notag\\
+ &a_2[-\i\alpha_+^{-1}, 1, 0, 0]\e{(\kappa_+ - \i m_x)x} \notag\\
+ &a_3[0, 0, \i\alpha_-, 1]\e{(\kappa_- + \i m_x)x} \notag\\
+ &a_4[0, 0, -\i\alpha_-^{-1}, 1]\e{-(\kappa_- - \i m_x)x} \Big),
\end{align}
where $a_j$ are the scattering coefficients and we have introduced:
\begin{align}
\kappa_\pm &= \sqrt{m_z^2 + (k_y \pm m_y)^2},\notag\\
\alpha_\pm &= -[\kappa_\pm - (k_y \pm m_y)]/m_z.
\end{align}
It is instructive to consider in some more detail why the wavefunction $\psi_F$ has the form of Eq. (\ref{eq:F}). Diagonalizing the Hamiltonian in the F region provides two eigenvalues for the electron-like quasiparticles $\varepsilon_\pm^e = \pm\sqrt{(k_x+m_x)^2+(k_y+m_y)^2+m_z^2}$, and two eigenvalues $\varepsilon_\pm^h = \pm\sqrt{(k_x-m_x)^2+(k_y-m_y)^2+m_z^2}$ for the hole-like quasiparticles. Since we are considering an incident electron with $\varepsilon>0$ from the N side, where the chemical potential is assumed to satisfy $\mu_N\gg\{\Delta_0,\varepsilon\}$, this excitation enters the F region in the band $\varepsilon_+^e = \sqrt{(k_x+m_x)^2+(k_y+m_y)^2+m_z^2}$. We illustrate this in Fig. \ref{fig:scattering_NF}. The Andreev-reflection process occurring at the F$\mid$S interface gives rise to a hole excitation by removing an electron from the band $\varepsilon_-^e$. The hole dispersion is opposite in sign to the electron dispersion of the band it was generated in, and therefore in this case follows from $\varepsilon_+^h$, which means that its group velocity is parallel to its momentum. In effect, this is therefore a specular Andreev reflection. By constructing the eigenvectors for the excitations in the bands $\varepsilon_+^e$ and $\varepsilon_+^h$, one finally arrives at Eq. (\ref{eq:F}) when taking into account both left- and right-moving exictations.

Finally, we write down the wavefunction in the superconducting (S) region of the surface which includes a contribution from both an electron-like and hole-like quasiparticle. As we shall see later, the superconducting spin-triplet case is qualitatively different from the spin-singlet case. \textit{The wavefunction below is therefore only valid for the spin-singlet case}, such as $s$-wave and $d$-wave, and reads:
\begin{align}
\psi_S = \e{\i k_y y} \Big(&t_e[\e{\i\beta}, \e{\i(\beta+\theta')}, -\e{\i(\theta'-\gamma_+)},\e{-\i\gamma_+}]\e{\i k_x'x} \notag\\
+ &t_h[1,-\e{-\i\theta'}, \e{\i(\beta-\theta'-\gamma_-)},\e{\i(\beta-\gamma_-)}]\e{-\i k_x'x}\Big).
\end{align}
Note that in writing down the above wavefunction we have taken into account the possibility of anisotropic pairing, and consequently defined 
\begin{align}
\e{\i\beta} = u_+/u_-,\; u_\pm = \sqrt{\frac{1}{2}(1 \pm \sqrt{\varepsilon^2-|\Delta(\theta')|^2}/\varepsilon)},\notag\\
\e{\i\gamma_\pm} = \Delta(\theta_\pm)/|\Delta(\theta_\pm)|,\; \theta_+=\theta', \theta_-=\pi-\theta'.
\end{align}
A difference in doping level between the N and S regions is accounted for by $\mu_N\sin\theta = \mu_S\sin\theta'$, since in an experimental situation the S region is often heavily doped $\mu_S\gg(\varepsilon,\Delta_0)$. The doping level can be controlled electrically by means of an applied gate voltage. If the N region is also doped away from the Dirac point, $\mu_N=\mu_S$ leads to $\theta'=\theta$. In the next section, we set $\mu_N=\mu_S \gg (\varepsilon,\Delta_0)$ unless specified otherwise.

In order to calculate the Andreev-bound state energies and the conductance spectra of this junction, we need to solve for the scattering coefficients. By matching the wavefunctions at each interface, i.e. $\psi_N = \psi_F$ at $x=0$ and $\psi_F =\psi_S$ at $x=L$, we obtain the system of equations $\mathcal{A}\mathbf{y} = \mathbf{b}$ where we have defined:
\begin{align}
\mathcal{A} &= \begin{pmatrix} 
\mathcal{A}_1 & \mathcal{A}_2\notag\\
\mathcal{A}_3 & \mathcal{A}_4 \notag\\
\end{pmatrix}, \notag\\
\mathbf{y} &= [r_e,r_h,a_1,a_2,a_3,a_4,t_e,t_h],\notag\\
\mathbf{b} &= [-1,-\e{\i\theta},0,0,0,0,0,0].
\end{align}
The analytical expressions for the matrices $\mathcal{A}_j$ can be found in the Appendix. After solving the above system for the unknown coefficients $\mathbf{y}$, the normalized conductance $G/G_0$ can be calculated according to the formula:
\begin{align}
G = \int^{\pi/2}_{-\pi/2} \text{d}\theta\cos\theta[1 + |r_h(\theta)|^2-|r_e(\theta)|^2],
\end{align}
and the normalization constant is chosen as $G_0=G(|eV|\gg\Delta_0)$ as usually done in experiments. We will also be interested in the bound-state energies of the junction, which correspond to resonant states that persist even in the limit of a vanishing normal-state conductance. This is modelled by letting the width $L$ of the junction become very large compared to all other length scales. These resonant states are found analytically by identifying the energies $\varepsilon$ where the probability for normal reflection vanishes, i.e. $r_e=0$. 

\subsubsection{$s$-wave singlet pairing}

\begin{figure}[t!]
\centering
\resizebox{0.48\textwidth}{!}{
\includegraphics{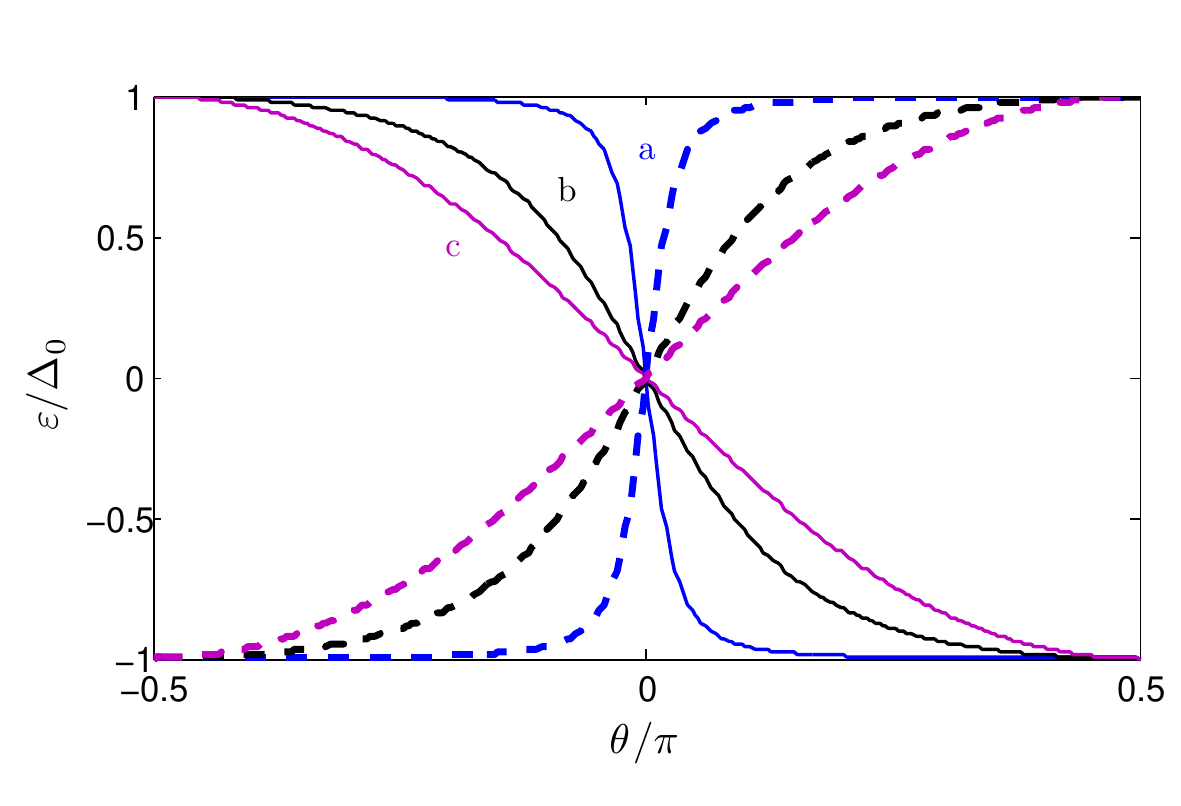}}
\caption{(Color online) Plot of the bound-state dispersion in the $s$-wave case for several values of $|m_z|/\mu$. The chirality of the bound-state is controlled by the sign of $m_z$. The solid lines correspond to $m_z>0$ while the dashed lines correspond to $m_z<0$. a: $|m_z|/\mu=0.1$, b: $|m_z|/\mu=0.5$, c: $|m_z|/\mu=0.9$. The superconducting gap $\Delta_0$ is used as the fundamental energy unit here, and we set $\mu/\Delta_0=100$.
}
\label{fig:boundswave} 
\end{figure}

For a spin-singlet symmetry one finds that 
\begin{align}
\underline{\Delta}(\vk) = \Delta(\vk)\i\underline{\sigma_y}.
\end{align}
The $s$-wave case has an isotropic order parameter $\Delta(\vk) = \Delta_0$, and diagonalization of Eq. (\ref{eq:H}) then yields the standard eigenvalues 
\begin{align}\label{eq:eigenswave}
\varepsilon = \eta\sqrt{(v_F|\vk|-\beta\mu)^2+|\Delta_0|^2},\; \eta=\pm1,\; \beta=\pm1.
\end{align}
Employing the strategy described in the previous section, we may calculate the resonant states by looking for energies that produce $r_e=0$ when $L\to \infty$. When $m_y=0$, one is able to write down a manageable analytical expression for the proper condition related to the formation of bound-states:
\begin{align}\label{eq:condswave}
\e{2\i\beta}\tau_+ + \tau_- &= 0.
\end{align}
Here, we have defined the auxiliary quantities:
\begin{align}\label{eq:quant}
\tau_\pm &= \sin\delta + \sin(2\theta+\delta) \pm[\sin\theta + \sin(2\delta+\theta)],\notag\\
\nu &= (k_y-\kappa)/m_z,\; \delta=-\i\ln(\nu/\i),\; \kappa = \sqrt{k_y^2+m_z^2}.
\end{align}
In the range $|\varepsilon|<\Delta_0$, one may write
\begin{align}
\beta = \text{atan}[\Delta_0\sqrt{1-(\varepsilon/\Delta_0)^2}/\varepsilon],
\end{align}
which upon insertion into Eq. (\ref{eq:condswave}) yields the following solution for the bound-state energy:
\begin{align}\label{eq:boundswave}
\varepsilon &= \Delta_0\text{sgn}\{\mathcal{C_-}\}/\sqrt{1 + \mathcal{C_-}^2},
\end{align}
where we have introduced:
\begin{align}
\mathcal{C_\pm} = \tan\Big[\text{ln}\Big(\pm \frac{\tau_-/\tau_+}{2\i}\Big)\Big].
\end{align}
We have checked analytically that Eq. (\ref{eq:boundswave}) is identical to Eq. (7) in Ref.~\onlinecite{tanaka_prl_09}. To explore how the magnetization influences the bound-state dispersion, we plot in Fig. \ref{fig:boundswave} the bound-state energy versus the angle of incidence (or equivalently the transverse momentum index) for several choices of $m_z$. It is seen that the bound-states have a dispersion only near $\theta=0$ when the Zeeman field is small, $|m_z|\ll\mu$. A zero-energy solution is seen to be allowed for normal incidence as long as $m_z$ is finite. It should also be noted that the chirality of the bound-states are determined by $\text{sgn}\{m_z\}$. To see this, note that $m_z\to(-m_z)$ leads to $\delta\to\delta+\pi$. Due to the symmetry relations $\tau_\pm(\delta+\pi) = -\tau_\mp(\delta)$ and ln$(\tau_-/\tau_+)$ = -ln$(\tau_+/\tau_-)$, it follows from the definition of $\mathcal{C_\pm}$ that 
\begin{align}
\text{sgn}\{\mathcal{C_\pm}\} \propto \text{sgn}\{m_z\}.
\end{align}

\begin{figure*}
\centering
\resizebox{0.95\textwidth}{!}{
\includegraphics{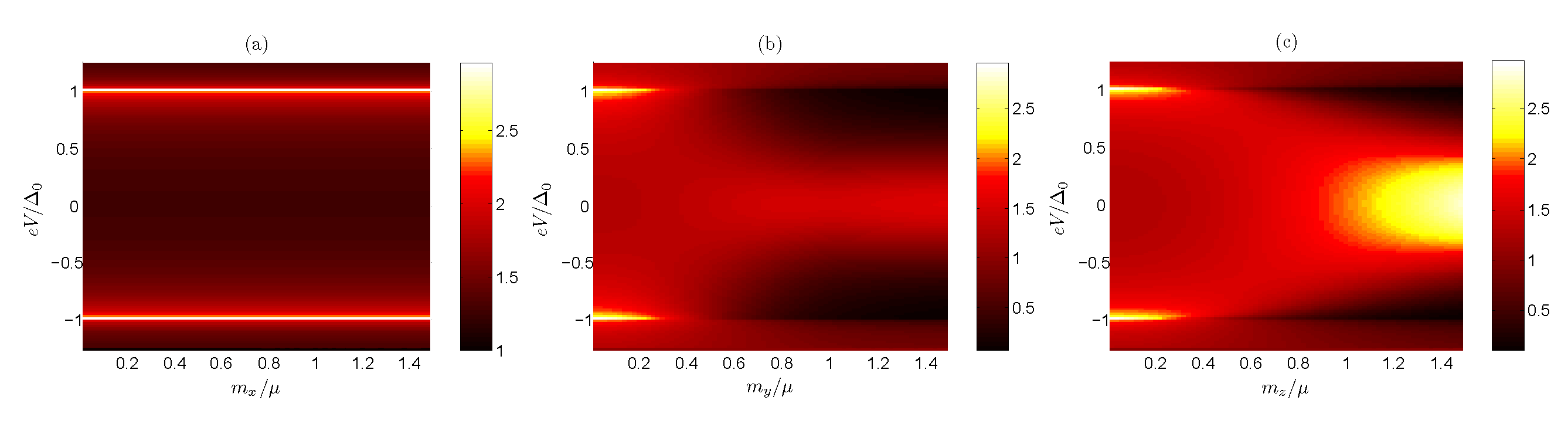}}
\caption{(Color online) Plot of the conductance spectra for an N$\mid$F$\mid$$s$-wave junction for a magnetization direction along the (a) $x$-axis, (b) $y$-axis, and (c) $z$-axis. We have set $\mu L = 1$ and $\mu/\Delta_0=100$.
}
\label{fig:condswave} 
\end{figure*}

To investigate how the presence of such bound-states is manifested in the experimentally accessible electrical conductance, we plot this quantity in Fig. \ref{fig:condswave}. This is done by solving numerically for the scattering coefficients, which enables us to consider also the case $m_y\neq0$. The dependence on the magnetization orientation is shown by considering a pure (a) $m_x$-component, (b) $m_y$-component, and (c) $m_z$-component. Interestingly, the dependence in all of these three cases are qualitatively very different, even though the $s$-wave order parameter is rotationally invariant. As we shall see in what follows, this stems from the unique band-structure of the surface-states of the topological insulator.

The only common feature all three plots in Fig. \ref{fig:condswave} have is the two coherence peaks at $\varepsilon=\Delta_0$ which exist when $m_j\to 0$, $j\in\{x,y,z\}$. In Fig. \ref{fig:condswave}(a), the conductance is invariant with respect to $m_x$, whereas in (b) it displays a considerable dependence on the magnitude of $m_y$. To explain this feature, the key point to observe is that the $k_y$-component of the momentum is conserved in the geometry under consideration since translational invariance holds in this direction. In this way, the component of the wavevector in the $x$-direction of propagation will in general contain a real and imaginary part, where the latter causes the wave to be evanescent (decaying). Even at normal incidence $\theta=0$, the wave becomes evanescent whenever $m_y\neq0$. In contrast, $m_x$ only influences the real part of the wavevector and does not influence the resistance of the junction since it is of no consequence for length scale of the decaying modes. Therefore, the conductance remains invariant under a change of $m_x$ whereas it is strongly dependent on the value of $m_y$. In fact, as $m_y$ increases the overall tendency of the conductance is that it is suppressed. In (c), we consider how the conductance changes when varying the "mass gap" component $m_z$. As $m_z$ grows, it is seen that the appearance of the bound state is manifested by a large enhancement of the zero-bias conductance. Therefore, the two finite-energy peaks are merged into one zero-energy resonance. Finally, we note that the $d_{x^2-y^2}$-wave case is qualitatively similar to the $s$-wave case.

\subsubsection{$p$-wave triplet pairing}

Turning to the spin-triplet case, the gap matrix now reads:
\begin{align}
\underline{\Delta}(\vk) = (\mathbf{d}_\vk\cdot\underline{\sigma})\i\underline{\sigma_y}.
\end{align}
Let us consider a general triplet state $\mathbf{d}_\vk = \Delta(\vk)\hat{\mathbf{z}}$. Diagonalizing Eq. (\ref{eq:H}) produces the following eigenvalues:
\begin{align}\label{eq:eigenpwave}
\varepsilon = \eta v_F|\vk| -\beta\sqrt{\mu^2+|\Delta(\vk)|^2}.
\end{align}
This equation is qualitatively different than Eq. (\ref{eq:eigenswave}) for the $s$-wave case. Namely, the superconducting order parameter now renormalizes the chemical potential and the \textit{excitations remain gapless}. From the dispersion Eq. (\ref{eq:eigenpwave}) follows several anomalous properties. By evaluating the corresponding wavefunction, one may conclude that Andreev reflection is strongly suppressed at the interface to a non-superconducting region since there is no gap in the charge excitation spectrum that can retroreflect a hole quasiparticle. Moreover, we have verified that for any triplet symmetry the anomalous dispersion Eq. (\ref{eq:eigenpwave}) is obtained. It also holds even if the Dirac-like $\underline{H_0}$ is replaced with a Rashba-like $\underline{H_0}$ as mentioned previously. Thus, the results for singlet and triplet pairing differ qualitatively in a fundamental way, as the excitations are gapped in the former case whereas they remain ungapped in the latter case. The structure of the eigenvalue in Eq. (\ref{eq:eigenpwave}) appears to be a direct result of the band-structure in the topological insulator, where the spin couples directly to momentum through the term $\underline{\sigma}\cdot\mathbf{k}$ in the Hamiltonian. Due to the fact that spin will be parallel to the momentum, it follows that pairing between equal spins (triplet pairing) at $\vk$ and $-\vk$ is not possible. This should be distinguished from the case of graphene, where the operator $\mathbf{\sigma}$ does not represent physical spin, but rather a pseudospin index related to the sublattices \cite{katsnelson_nphys_06, uchoa_prl_07, linder_prl_08}.

\subsubsection{$d$-wave singlet pairing}

\begin{figure}[t!]
\centering
\resizebox{0.48\textwidth}{!}{
\includegraphics{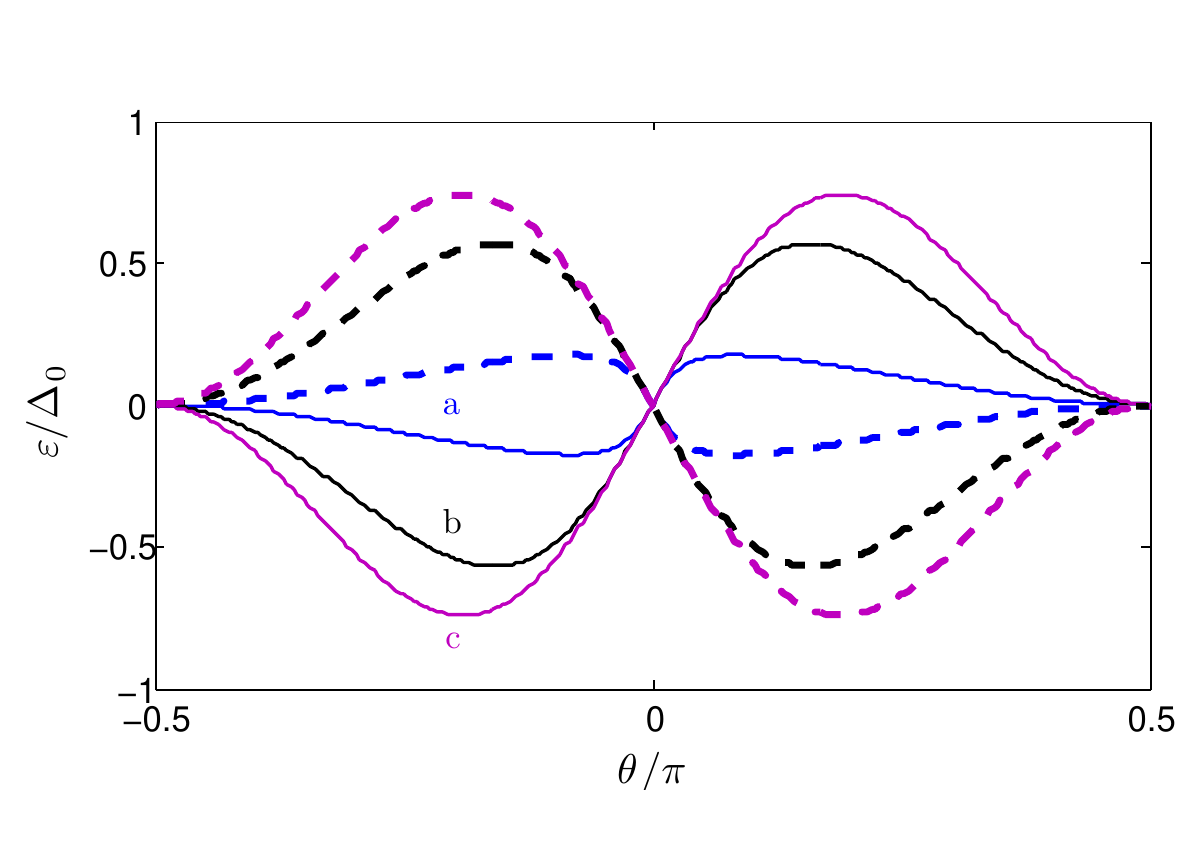}}
\caption{(Color online) Plot of the bound-state dispersion in the $d_{xy}$-wave case for several values of $|m_z|/\mu$. Similarly to the $s$-wave case, the chirality of the bound-state is controlled by the sign of $m_z$. The solid lines correspond to $m_z>0$ while the dashed lines correspond to $m_z<0$. a: $|m_z|/\mu=0.1$, b: $|m_z|/\mu=0.5$, c: $|m_z|/\mu=0.9$. We have set $\mu/\Delta_0=100$.
}
\label{fig:bounddwave} 
\end{figure}

We finally address the $d$-wave pairing state, focusing here on a $d_{xy}$-wave symmetry. The reason for why this symmetry is the most interesting is that it is known to produce zero-energy surface states in the high-$T_c$ cuprates. The order parameter may in this case be written as $\Delta(\vk) = \Delta(\theta) =  \Delta_0\cos(2\theta-\pi/2)$, and diagonalization of Eq. (\ref{eq:H}) then yields the standard eigenvalues 
\begin{align}
\varepsilon = \eta\sqrt{(v_F|\vk|-\beta\mu)^2+|\Delta(\theta)|^2},\; \eta=\pm1,\; \beta=\pm1.
\end{align}
Proceeding in the same fashion as the previously considered $s$-wave case, we find the following condition for the bound-state energies when $m_y=0$:
\begin{align}\label{eq:conddwave}
\e{2\i\beta}\tau_+ - \tau_- &= 0.
\end{align}
As seen, the only difference from Eq. (\ref{eq:condswave}) is the sign of the last term in the equation, although we shall see that this sign change has fundamental consequences. Eq. (\ref{eq:conddwave}) yields the following solution for the bound-state energy:
\begin{align}\label{eq:bounddwave}
\varepsilon &= |\Delta(\theta)|\text{sgn}\{\mathcal{C_+}\}/\sqrt{1 + \mathcal{C_+}^2}.
\end{align}
We now consider how the magnetization influences the bound-state dispersion in Fig. \ref{fig:bounddwave}. The dispersion is weak, and the bound-state energy is close to $\varepsilon=0$, for $|m_z| \ll \mu$. Increasing $m_z$ ($m_z>0$) in the $d$-wave case has the important effect of accomodating finite-energy bound-states when moving away from normal incidence. It is noted that in the same way as for the $s$-wave case, the chirality of the bound-state is determined by the sign of $m_z$.

\begin{figure*}
\centering
\resizebox{0.95\textwidth}{!}{
\includegraphics{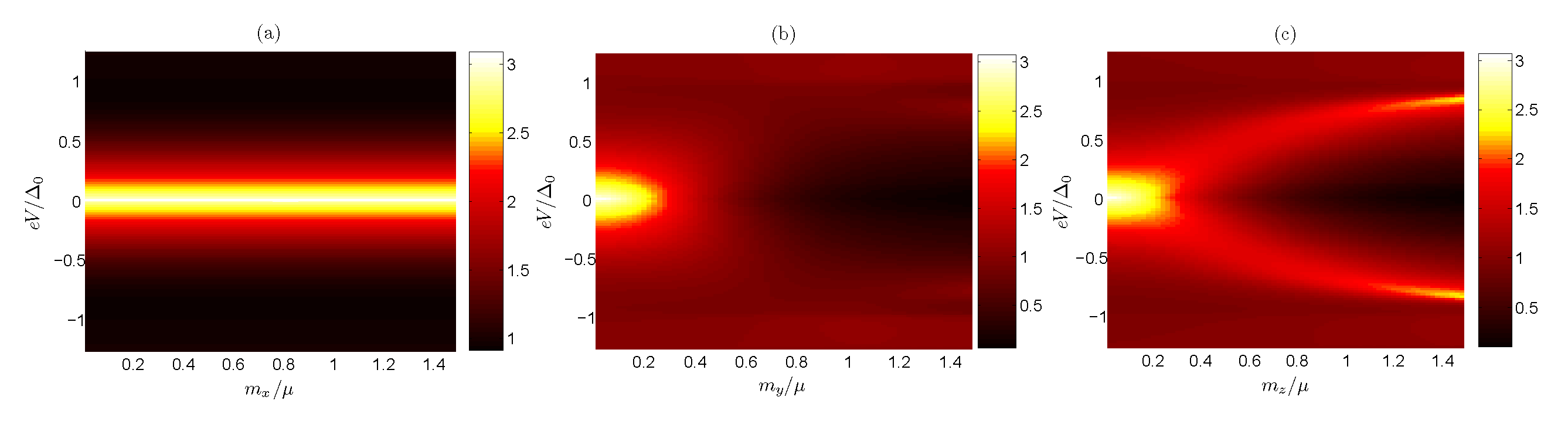}}
\caption{(Color online) Plot of the conductance spectra for an N$\mid$F$\mid$$d_{xy}$-wave junction for a magnetization direction along the (a) $x$-axis, (b) $y$-axis, and (c) $z$-axis. We have set $\mu L = 1$ and $\mu/\Delta_0=100$.
}
\label{fig:conddwave} 
\end{figure*}

The zero-energy states appear in the $d_{xy}$-wave case even in the absence of magnetization, as may be shown by solving the scattering problem for a N$\mid$$d_{xy}$-wave junction on a topological insulator when there is a Fermi-surface mismatch between the N and S region. We now show that these zero-energy bound-states are Majorana fermions, in contrast to the zero-energy states realized in the topologically trivial high-$T_c$ cuprates. The essential point in this context is the spin-degeneracy of the Fermi surface in the latter case, whereas for a topological insulator this degeneracy is lifted. In both cases, the $4\times4$ BdG Hamiltonian $\hat{H}$ satisfies a particle-hole symmetry 
\begin{align}
\Theta\hat{H}(\vk)\Theta = -\hat{H}^*(-\vk),
\end{align}
where we have introduced the matrix \cite{sato_prb_09}
\begin{align}
\Theta = \begin{pmatrix} \underline{0} & \underline{1}\\ \underline{1} & \underline{0} \\ \end{pmatrix}.
\end{align}
From this property, one may prove that if 
\begin{align}
\psi_\varepsilon = [u_1(\vk),u_2(\vk),v_1(\vk),v_2(\vk)]
\end{align}
is an eigenfunction for the eigenvalue $\varepsilon$, then 
\begin{align}
\Theta\psi_\varepsilon(-\vk)^* &= \psi_{-\varepsilon}(\vk)\notag\\
&= [v_1^*(-\vk),v_2^*(-\vk),u_1^*(-\vk),u_2^*(-\vk)]
\end{align}
is an eigenfunction for $(-\varepsilon)$. For a zero-energy bound state $\varepsilon=0$, one must have $\psi_\varepsilon = \psi_{-\varepsilon}$, leading to internal symmetry relations between the coherence factors such as $u_1(\vk)=v_1^*(-\vk)$. The Bogoliubov quasiparticle creation operator for this state is constructed in the usual way as \begin{align}
\gamma^\dag(\vk) &= u_1(\vk) c_\uparrow \dag(\vk) + u_2(\vk) c_\downarrow^\dag(\vk) \notag\\
&+ v_1(\vk) c_\uparrow(-\vk) + v_2(\vk) c_\downarrow(-\vk)
\end{align}
Thus, we see that the Majorana criterion $\gamma(\vk) = \gamma^\dag(-\vk)$ is satisfied. Now, the distinction between the zero-energy state in the cuprates and the present context of a topological insulator is precisely the spin-degeneracy which allows one to split up the $4\times4$ BdG equations to two separate $2\times2$ equations per spin. Due to the band-structure on the surface of a topological insulator, the $\varepsilon=0$ solution is not spin-degenerate and we obtain only one zero-energy mode. As pointed out in Ref. \cite{sato_prb_09}, this guarantees the Majorana nature of the fermion. We reemphasize that this is different from topologically trivial N$\mid$$d_{xy}$-wave junctions, formed \eg by a normal metal contacted to yttrium barium copper oxide (YBCO), where the zero-energy solutions are spin-degenerate.

We now investigate how the presence of such bound-states are manifested in an experimentally accessible quantity, namely the electric conductance. To do so, we consider an N$\mid$F$\mid$$d$-wave junction to check how the magnetization can be used to manipulate the transport properties and give signatures of the surface-states. In Fig. \ref{fig:conddwave}, we plot the conductance, which normally is expected to produce the well-known zero-bias conductance peak (ZBCP) due to midgap resonant states \cite{hu_prl_94, tanaka_prl_95}.

The common feature for all magnetization directions in (a)-(c) is that a zero-bias peak is present when $m_j\to0$, in agreement with our previous analytical finding. Due to the coupling between spin and momentum in the band structure of the surface of a topological insulator, it is interesting to check whether the direction of the magnetization influences the conductance spectra. In a topologically trivial N$\mid$F$\mid$$d_{xy}$-wave junction, one can prove analytically that the conductance is invariant with respect to the direction of the magnetization $\mathbf{m}$ of the F layer. Increasing the exchange field in the F region, the ZBCP splits in the conventional case \cite{kashiwaya_prb_99}, similarly to Fig. \ref{fig:conddwave}(c). We here show that in complete contrast to the topologically trivial case, the conductance now features a strong dependence on the magnetization orientation. We consider a magnetization in the $\hat{\mathbf{x}}$- and $\hat{\mathbf{y}}$-direction in Fig. \ref{fig:conddwave}(a) and (b), respectively. It is seen that depending on the magnetization orientation, the conductance features three qualitatively different types of behavior. For $\mathbf{m}\parallel\hat{\mathbf{x}}$, $G/G_0$ is invariant upon increasing $m_x$. For $\mathbf{m}\parallel\hat{\mathbf{y}}$, the ZBCP vanishes upon increasing $m_y$. For $\mathbf{m}\parallel\hat{\mathbf{z}}$, the ZBCP is split upon increasing $m_z$. In fact, the evolution of the conductance spectra in (c) is opposite to the $s$-wave case upon increasing the magnetization: the zero-bias peak is split into two finite-energy resonances. In effect, this means that the characteristic features in the conductance spectra of $s$-wave and $d$-wave superconductors can be completely reversed by introducing a Zeeman field in the topological insulator.
The strong sensitivity to the direction of $\mathbf{m}$ is a new feature compared the topologically trivial case which pertains directly to the anomalous band-structure of the topological insulator. The difference between the $m_x$ and $m_y$ cases shown in (a) and (b) is explained in the same way as for the $s$-wave pairing scenario. As seen, an increase in $m_y$ eventually suppresses the influence of superconductivity and the conductance is reduced.

\subsubsection{Singlet-Triplet mixing}\label{sec:mix}

Due to the lack of inversion symmetry and concomitant presence of asymmetric spin-orbit coupling in the topological insulator, one might envision a mixed $s+p$-wave superconducting state induced by the proximity effect. For a chiral $p$-wave state $\Delta(\vk) = \Delta_p\e{\i\theta}$ ($\mathbf{d}_\vk$-vector along the $\mathbf{\hat{z}}$-axis), the resulting eigenvalues read 
\begin{align}
\varepsilon &= \eta\sqrt{v_F|\vk|^2+\mu^2+\Delta_s^2+\Delta_p^2 -2\beta\sqrt{R}},\notag\\
R &= v_F^2|\vk|^2(\mu^2+\Delta_p^2) + \Delta_s^2\Delta_p^2\cos^2 2\theta,\; \eta=\pm1,\; \beta=\pm1.
\end{align}
In the limits $\Delta_s\to0$ and $\Delta_p\to0$, this reduces to the previous expressions in this work. As seen, the 
presence of an $s$-wave component ensures that the spectrum is gapped, while a coupling between the $s$-wave and $p$-wave components render the excitations sensitive to the angle of incidence $\theta$. The analytical expression for the corresponding wavefunction in the $s+p$-wave case is unwieldy, and we defer from any further treatment of the transport properties of such a state here. The key point we wish to illustrate with this discussion is that the $s$-wave component (in general, the spin-singlet component) is necessary to obtain a gapped spectrum, but the interplay between spin-singlet and spin-triplet pairing nevertheless gives rise to a quasiparticle spectrum which is sensitive to the direction of propagation.

\subsection{Josephson current}\label{sec:jos}

We now turn to a study of the Josephson current in an S$\mid$F$\mid$S structure deposited on top of the topological insulator. Since the F region is assumed to be insulating, such as EuO or EuS with band gaps of a few eV, the transport is ensured to take place through Andreev bound-states formed on the surface of the topological insulator. These bound-state energies can be obtained by matching the wavefunctions in a similar way as in the previous section. The wavefunction in the F region remains the same, while we relabel $\psi_S \to \psi_S^{r}$ and concomitantly $\{t_e,t_h\} \to \{t_e^{r}, t_h^{r}\}$ where the superscript 'r' stands for the right region $x>L$. It then remains to specify the wavefunction in the left superconducting region, which reads:
\begin{align}
\psi_S^{l} = \e{\i k_y y} \Big(&t_e^{l}[\e{\i\beta}, -\e{\i(\beta-\theta')}, \e{\i(-\theta'-\gamma_-)},\e{-\i\gamma_-}]\e{-\i k_x'x} \notag\\
+ &t_h^{l}[1,\e{\i\theta'}, -\e{\i(\beta+\theta'-\gamma_+)},\e{\i(\beta-\gamma_+)}\e{\i k_x'x}\Big).
\end{align}
To identify the energy resonances for this system, we look for an energy eigenvalue $\varepsilon$ which gives a non-trivial solution for the boundary conditions $\psi_S^l = \psi_F$ at $x=0$ and $\psi_F = \psi_S^r$ at $x=L$. It can be directly verified by setting up these equations that this eigenvalue must satisfy $\text{det}\mathcal{M} = 0$, where
\begin{align}
\mathcal{M} &= \begin{pmatrix}
\mathcal{M}_1 & \mathcal{M}_2 \\
\mathcal{M}_3 & \mathcal{M}_4 \\
\end{pmatrix},
\end{align}
and the $2\times2$-matrices $\mathcal{M}_j$ are given explicitly in the Appendix. After some cumbersome but straight-forward algebra, we obtain the following expression for the bound-state energy:
\begin{align}\label{eq:boundjos}
\varepsilon = |\Delta(\theta)|\sqrt{\frac{1}{2}\Big(1 - \frac{\Gamma(\Delta\phi)}{\zeta}\Big)}.
\end{align}
where we have defined
\begin{align}
\Gamma(\Delta\phi) &= 2\cos(\Delta\phi-2m_xL)\cos^2\theta\sin^2\delta \notag\\
&- \sigma(\cos^2\theta+\cos^2\delta)[1-\cosh(2\kappa L)],\notag\\
\zeta &= 2\cos^2\theta\cos^2\delta - \cos^2\theta-\cos^2\delta\notag\\
&+\cosh(2\kappa L)(\cos^2\delta-\cos^2\theta).
\end{align}
The definition of quantities such as $\kappa$ and $\delta$ are found in Eq. (\ref{eq:quant}), and we have $\sigma=-1$ in the $s$-wave case while $\sigma=1$ in the $d_{xy}$-wave case. Above, we have set $m_y=0$ since an analytical expression for the bound-state becomes unwieldy otherwise. The normalized Josephson current may now be evaluated according to the standard expression:
\begin{align}
I/I_0 &= \int^{\pi/2}_{-\pi/2} \text{d}\theta \cos\theta\tanh(\beta\varepsilon/2)\frac{\text{d}\varepsilon}{\text{d}\Delta\phi},
\end{align}
which upon insertion of Eq. (\ref{eq:boundjos}) produces:
\begin{align}\label{eq:jos}
I/I_0 &=  \sin(\Delta\phi-2m_x L)\int^{\pi/2}_{-\pi/2} \text{d}\theta \frac{\Delta_0[g(\theta)]^2\cos^3\theta\sin^2\delta}{\varepsilon \zeta [\tanh(\beta\varepsilon/2)]^{-1}}.
\end{align}
Above, we have defined $\Delta(\theta) = \Delta_0g(\theta)$, such that $g(\theta)=1$ for $s$-wave pairing and $g(\theta)=\cos(2\theta-\pi/2)$ for $d_{xy}$-wave pairing. Also, $\Delta\phi$ denotes the superconducting phase difference. Below, we shall consider both the current-phase relation and the dependence of the critical current on the temperature $T$ and the junction width $L$. The critical current is measured experimentally as 
\begin{align}
I_c/I_0 = \text{max}_{\Delta\phi}(I/I_0).
\end{align}
To make contact with realistic experimental parameters, we estimate the Fermi velocity as \cite{zhang_nphys_09} $v_F\simeq5\times10^{5}$ m/s. Due to the lattice mismatch between the host proximity materials and the topological insulator, the induced superconducting order parameter $\Delta_0$ can be expected to be substantially reduced in magnitude on the surface of the topological insulator, and may be assumed to satisfy $\Delta_0 \simeq 0.5$ meV in the $d$-wave case. The superconducting coherence length is then $\xi \simeq 650$ nm. For a standard $s$-wave superconductor such as Al or Nb, the gap is typically much smaller than in a high-$T_c$ cuprate, and here one might estimate $\Delta_0\simeq0.1$ meV. To ensure ballistic transport, shorter junction are preferable since $L$ must then be smaller than the mean free path $l_\text{mfp}$. Also, the formula for the Josephson current above is valid under the condition that the junction length satifies $L\ll\xi$. 
We underline that by considering Eq. (\ref{eq:jos}), we may immediately infer that the current no longer necessarily vanishes at $\Delta\phi=\{0,\pi\}$ as long as $m_x\neq0$. Therefore, it is possible to actively tune the current-phase relation by means of the transverse magnetization component $m_x$ which enters in the gauge invariant phase-difference between the superconducting order parameters. 

\begin{figure}[t!]
\centering
\resizebox{0.48\textwidth}{!}{
\includegraphics{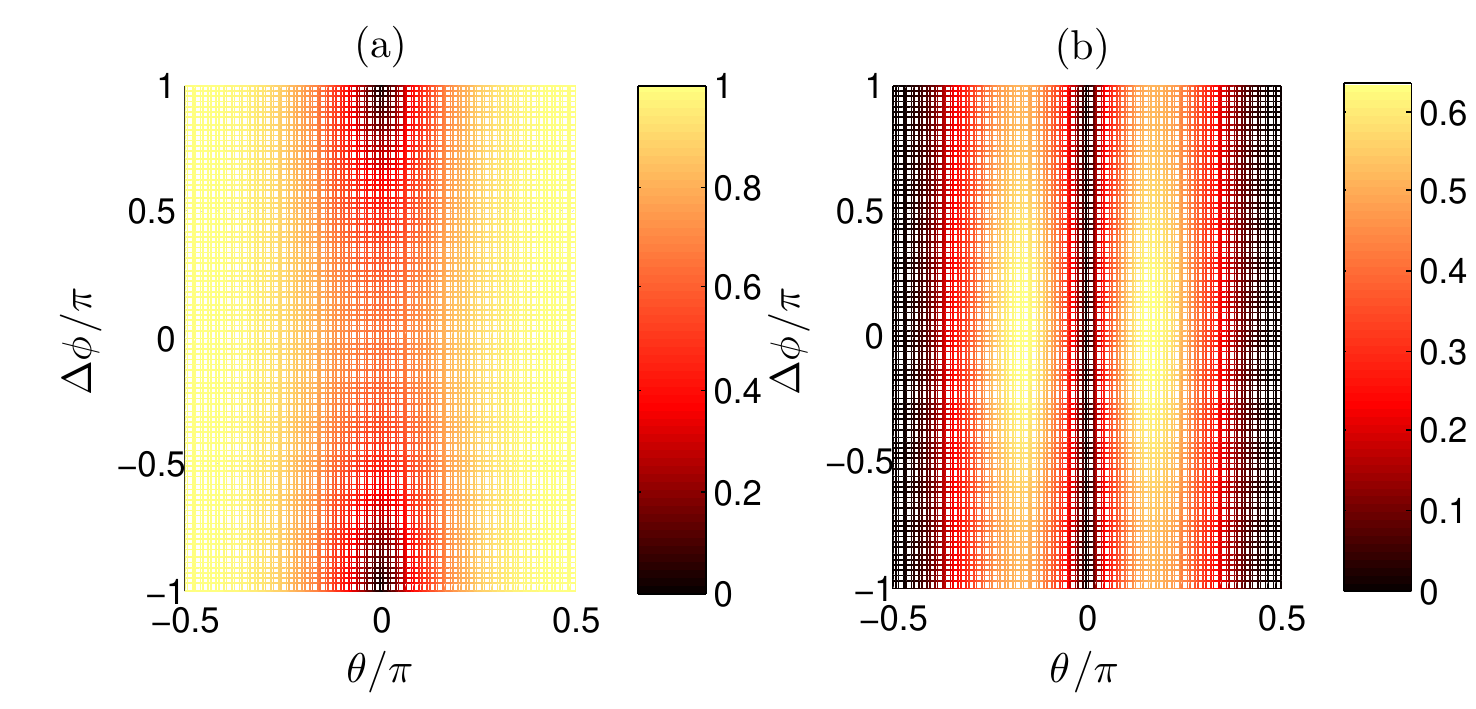}}
\caption{(Color online) Plot of the bound-state energy in the Josephson S$\mid$F$\mid$S junction with an (a) $s$-wave and (b) $d_{xy}$-wave symmetry. We have set $\mu/\Delta_0=100$, $m_z/\mu=0.5$, and $L/\xi=0.02$. Since $m_x$ simply corresponds to a shift in $\Delta\phi$, we have set $m_x=0$ here.
}
\label{fig:josephson_bound} 
\end{figure}

\begin{figure}[t!]
\centering
\resizebox{0.48\textwidth}{!}{
\includegraphics{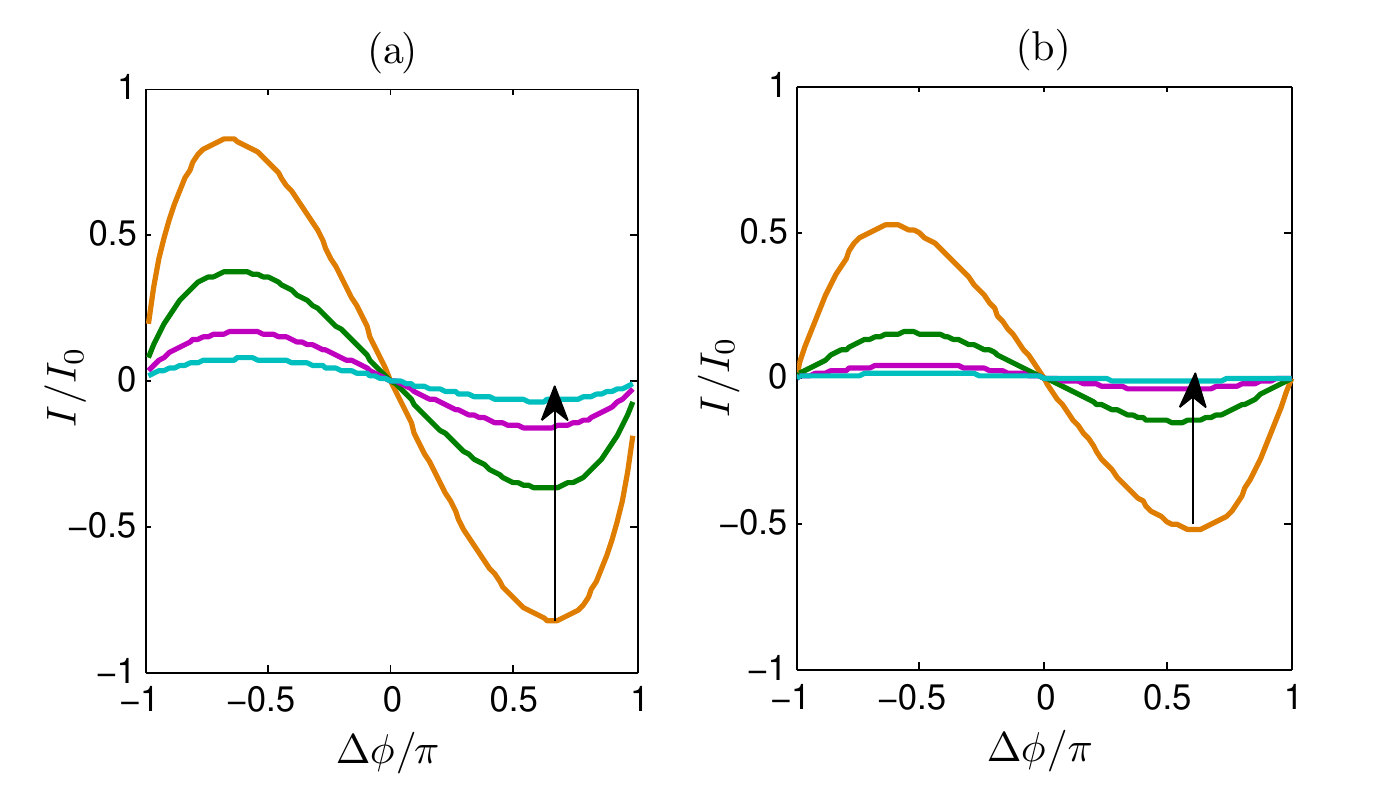}}
\caption{(Color online) Plot of the current-phase relation in the Josephson S$\mid$F$\mid$S junction with an (a) $s$-wave and (b) $d_{xy}$-wave symmetry. We have set $\mu/\Delta_0=100$, $m_z/\mu=0.5$, and considered the values $L/\xi=0.01, 0.02, 0.03, 0.04$ in the direction of the arrow. Since $m_x$ simply corresponds to a shift in $\Delta\phi$, we have set $m_x=0$ here.
}
\label{fig:josephson_phase} 
\end{figure}

In Fig. \ref{fig:josephson_bound}, we consider first the energy of the Andreev-bound state in the S$\mid$F$\mid$S junction, setting for simplicity $m_x=0$ since a non-zero $m_x$ simply would correspond to a constant shift of $\Delta\phi$. As mentioned previously, we have set $m_y=0$ for all results pertaining to the Josephson current to obtain analytical expressions. For normal incidence, it is seen that the bound-state energy goes to zero at $\Delta\phi=\{-\pi,\pi\}$ for the $s$-wave case, whereas it is maximal ($\varepsilon=\Delta_0$) when $\theta\to\pm\pi/2$. This is in agreement with the finding of Ref.~\onlinecite{tanaka_prl_09}. In the $d_{xy}$-wave case shown in Fig. \ref{fig:josephson_bound}(b), the angular dependence of the bound-state energy is modified strongly due to the anisotropy of the gap. Experimentally, one may probe the Andreev-bound states indirectly through their influence on the current-phase relation, which we consider in Fig. (\ref{fig:josephson_phase}). The phase difference can be actively manipulated either by means of current-biasing the junction or tuning the magnetization component $m_x$. The curves are translated to left or right for non-zero $m_x$, depending on its sign \footnote{There is a typo in the figure caption of Fig. 4 in Ref.~\onlinecite{tanaka_prl_09}: curve a is plotted for $m_x/m_z=0$ and not $m_x/m_z=1$.}. As seen, the current-phase relation is qualitatively similar for the $s$-wave and $d_{xy}$-wave cases, although its magnitude is reduced for the latter. This is related to the effective weakening of the gap upon Fermi surface averaging of its absolute value compared to the $s$-wave case. It is clear from Fig. \ref{fig:josephson_phase} that the phase-relation is not purely sinusoidal, but contains a contribution from higher harmonics.

\begin{figure}[t!]
\centering
\resizebox{0.48\textwidth}{!}{
\includegraphics{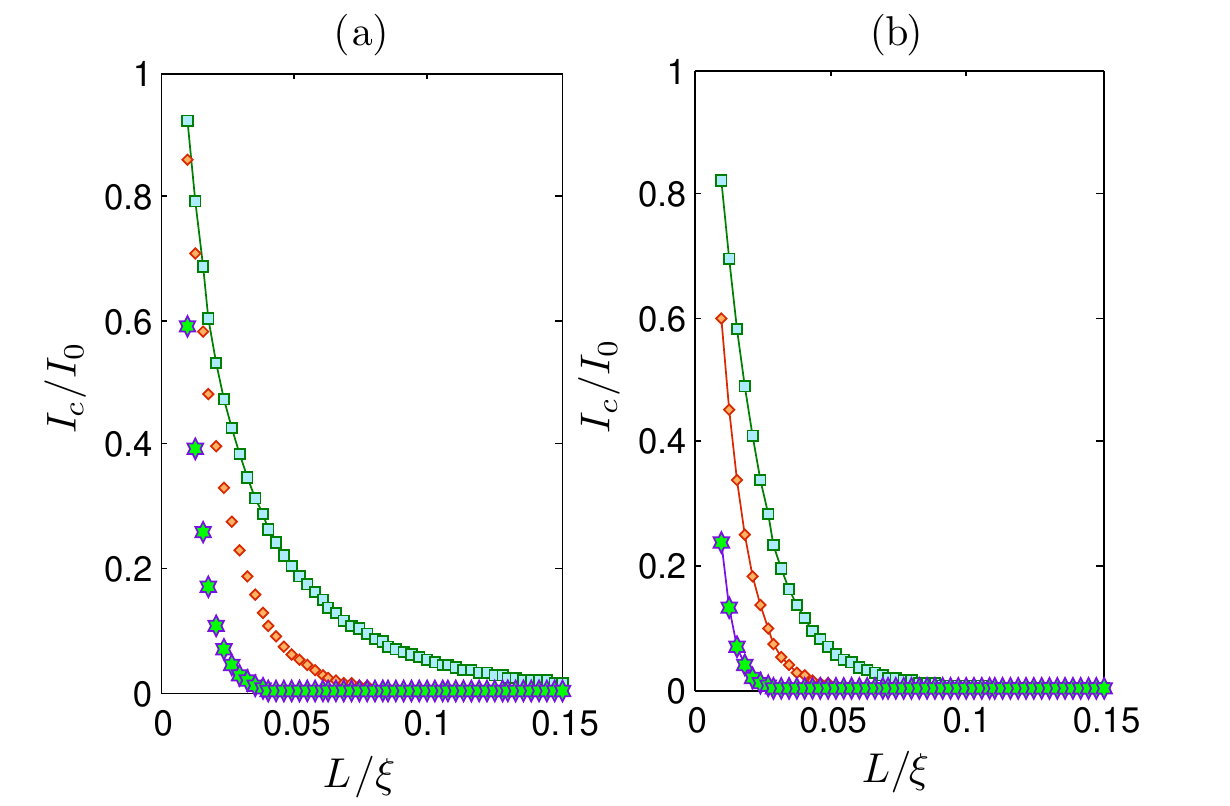}}
\caption{(Color online) Plot of the length-dependence of the critical current of the Josephson S$\mid$F$\mid$S junction with $T=0$ and an (a) $s$-wave and (b) $d_{xy}$-wave symmetry. We have set $\mu/\Delta_0=100$ and considered $m_z/\mu= 0.15, 0.40, 1.0$ from top to bottom.
}
\label{fig:josephson_length} 
\end{figure}

Next, we consider the dependence of the critical current on the length $L$ of the junction. This quantity is also routinely measured in the context of S$\mid$F$\mid$S junctions. The most notable feature is that there are no 0-$\pi$ oscillations in the current, in spite of the presence of an exchange field $m_z$ in the F region. To explain this, one should note that the influence of the exchange field $m_z$ is fundamentally different in the present scenario where we consider states residing on the surface of a topological insulator as compared to a normal metal. In the present case it induces a gap in the spectrum whereas in the latter case it splits the energy-bands of the majority- and minority-spins. As a result, the exchange field does not induce any finite center-of-mass momentum for the Cooper-pair and hence a monotonous decay of the critical current versus $L$ should be observed regardless of the magnitude of $\mathbf{m}$. The main difference between the $s$-wave and $d_{xy}$-wave case in Fig. \ref{fig:josephson_length} is that the current becomes suppressed more rapidly as $L$ grows in the latter scenario.

\begin{figure}[t!]
\centering
\resizebox{0.48\textwidth}{!}{
\includegraphics{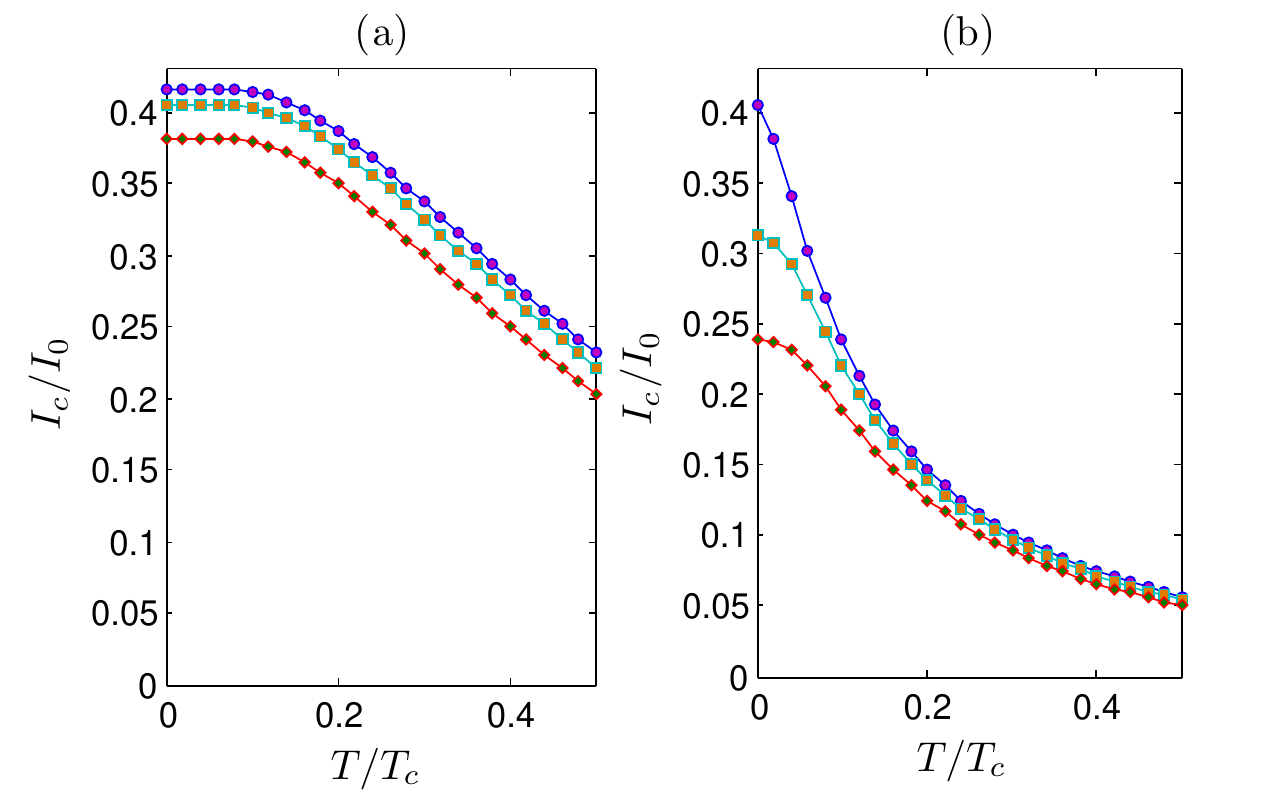}}
\caption{(Color online) Plot of the temperature-dependence of the critical current of the Josephson S$\mid$F$\mid$S junction with $L/\xi=0.03$ and an (a) $s$-wave and (b) $d_{xy}$-wave symmetry. We have set $\mu/\Delta_0=100$ and considered $m_z/\mu=0.02, 0.08, 0.14$ from top to bottom. As $m_z/\mu\to 0$, the junction becomes an effective S$\mid$N$\mid$S system.
}
\label{fig:josephson_temp} 
\end{figure}

It is natural to next address the question: what happens when the magnitude of the exchange field $m_z$ decreases and eventually vanishes? This would in effect render the junction into an S$\mid$N$\mid$S system. Several previous works \cite{tanaka_prb_97, tanaka_jpsj_00} have investigated a similar scenario when using a $d$-wave model relevant for the high-$T_c$ cuprates. In such a case, it was shown that the critical current becomes strongly enhanced at low temperatures in the $d_{xy}$-wave case, whereas it saturates in the $s$-wave case. The reason for this is the existence of zero-energy Andreev-levels in the $d_{xy}$-wave junction. Above, we considered the case $m_z\gg\Delta_0$ where such states are shifted to finite energies as indicated by our previous results for the conductance spectra. Now, we study how the temperature-dependence of the critical current evolves when $m_z$ decreases and compare the $s$-wave and $d_{xy}$-wave scenarios. This is shown in Fig. \ref{fig:josephson_temp}. In contrast to Fig. \ref{fig:josephson_length}, the current now behaves qualitatively different in (a) and (b) corresponding to the $s$-wave and $d_{xy}$-wave case. In (a), the current saturates at a constant value as $T/T_c\to0$, whereas in (b) there is a strong enhancement in the same regime which is more pronounced the smaller $m_z/\mu$ becomes. Thus, such an anomalous temperature-dependence serves as a signature for the zero-energy states on the topological surface in the same way as it does for the cuprates, and can be probed experimentally. We have verified that when $m_z\gg\Delta_0$, the temperature-dependence of the current is qualitatively the same in the $s$-wave and $d_{xy}$-wave cases.

We note that an inclusion of the orbital effect due to the vector potential $\mathbf{A}$ simply would add a component to the magnetization vector as a result of the linear energy-momentum dispersion. The predicted results in this work can be tested experimentally by fabricating a hybrid structure such as the one shown in Fig. \ref{fig:model}. In terms of actual materials, EuO or EuS might be suitable as ferromagnetic insulators in this context \cite{tedrow_prl_86}. For the $d$-wave superconductor, a high-$T_c$ cuprate such as YBCO would be appropriate. Finally, we also point out that the suppression of the superconducting order parameter near the interface region has not been taken into account here. Such an approximation is valid when there is a strong Fermi-surface mismatch, as considered throughout this paper. Nevertheless, it could be interesting to see if the gap suppression is able to host additional bound-states near the interface by employing a self-consistent solution of the order parameter. We leave such issues for future work.

\section{Conclusion}

In summary, we have considered the interplay between magnetic order and unconventional superconducting pairing on the surface of a topological insulator. We find that the charge excitation spectrum is rendered gapless for any spin-triplet state, such that bound-states are absent and Andreev reflection is strongly suppressed. For spin-singlet pairing, we find that the zero-energy surface states in the $d_{xy}$-wave case are now Majorana fermions, in contrast to the case of the topologically trivial high-$T_c$ cuprates. We have studied how Andreev-bound states and Majorana fermions are influenced by the internal phase of the superconducting order parameter, and find that the ZBCP being the hallmark of the $d_{xy}$-wave state is qualitatively strongly modified in the present context. In particular, it is highly sensitive to the magnetization orientation, in contrast to the topologically trivial case. Our findings can be directly tested through tunneling spectroscopy measurements, and we have estimated the magnitude of the necessary experimental quantities.

\acknowledgments

B.-S. Skagerstam is thanked for useful discussions. J.L. and A.S. were supported by the Research Council of Norway, 
Grants No. 158518/431 and No. 158547/431 (NANOMAT), and Grant No. 167498/V30 (STORFORSK). 

\begin{widetext}

\appendix

\section{Matrices $\mathcal{A}_j$ and $\mathcal{M}_j$}

Introducing the quantity $\delta_\pm = -\i\text{ln}(\alpha_\pm/\i)$, we may write
\begin{align}
\mathcal{A}_1 = \begin{pmatrix}
1 & 0 & \e{\i\delta_+} & \e{-\i\delta_+} \\
-\e{-\i\theta} & 0 & -1 & -1 \\
0 & 1 & 0 & 0 \\
0 & -\e{-\i\theta} & 0 & 0 \\
\end{pmatrix}&,\; \mathcal{A}_2 = \begin{pmatrix}
0 & 0 & 0 & 0 \\
0 & 0 & 0 & 0 \\
\e{\i\delta_-} & \e{-\i\delta_-} & 0 & 0 \\
-1 & -1 & 0 & 0 \\
\end{pmatrix},\; \mathcal{A}_3 = \begin{pmatrix}
0 & 0 & -\e{\i\delta_+ - \kappa_+L-\i m_xL} & -\e{-\i\delta_+ + \kappa_+L-\i m_xL} \\
0 & 0 & \e{-\kappa_+L - \i m_xL} & \e{\kappa_+ L -\i m_xL} \\
0 & 0 & 0 & 0 \\
0 & 0 & 0 & 0 \\
\end{pmatrix},\notag\\
\mathcal{A}_4 = &\begin{pmatrix}
0 & 0 & -\e{\i\beta} & -1 \\
0 & 0 & -\e{\i(\beta+\theta)} & \e{-\i\theta} \\
-\e{\i\delta_-+\kappa_- L +\i m_xL} & -\e{-\i\delta_- - \kappa_- L +\i m_xL} & \e{\i(\theta-\gamma_+)} & -\e{\i(\beta-\theta-\gamma_-)} \\
\e{\kappa_- L+\i m_xL} & \e{-\kappa_- L +\i m_xL} & -\e{-\i\gamma_+} & -\e{\i(\beta-\gamma_-)}\\
\end{pmatrix},
\end{align}
which are used to calculate the bound-states and conductance spectra in the N$\mid$F$\mid$S junction. To obtain the Josephson current in the S$\mid$F$\mid$S case, we use:
\begin{align}
\mathcal{M}_1 = &\begin{pmatrix}
\e{\i\beta} & 1 & \e{\i\delta_+}  & \e{-\i\delta_+} \\
-\e{\i(\beta-\theta')} & \e{\i\theta'} & -1 & -1 \\
\e{-\i(\theta'+\gamma_-+\phi/2)} & -\e{\i(\beta+\theta'-\gamma_+-\phi/2)} & 0 & 0 \\
\e{-\i(\gamma_-+\phi/2)} & \e{\i(\beta-\gamma_+ -\phi/2)} & 0 & 0 \\
\end{pmatrix},\; \mathcal{M}_2 = \begin{pmatrix}
0 & 0 & 0 & 0 \\
0 & 0 & 0 & 0 \\
\e{\i\delta_-} & \e{-\i\delta_-} & 0 & 0 \\
-1 & -1 & 0 & 0 \\
\end{pmatrix},\notag\\
\mathcal{M}_3 = &\begin{pmatrix}
0 & 0 & -\e{\i\delta_+-(\kappa_+ + \i m_x)L} & -\e{-\i\delta_+ + (\kappa_+ -\i m_x)L}\\
0 & 0 & \e{-(\kappa_+ + \i m_x)L} & \e{(\kappa_+ -\i m_x)L} \\
0 & 0 & 0 & 0 \\
0 & 0 & 0 & 0 \\
\end{pmatrix},\notag\\
\mathcal{M}_4 = &\begin{pmatrix}
0 & 0 & -\e{\i\beta} & -1 \\
0 & 0 & -\e{\i(\beta+\theta')} & \e{-\i\theta'} \\
-\e{\i\delta_- + (\kappa_- + \i m_x)L} & -\e{-\i\delta_- -(\kappa_- -\i m_x)L} & \e{\i(\theta'-\gamma_++\phi/2)} & -\e{\i(\beta-\theta'-\gamma_- +\phi/2)} \\
\e{(\kappa_-+\i m_x)L} & \e{-(\kappa_--\i m_x)L} & -\e{-\i\gamma_++\i\phi/2} & -\e{\i(\beta-\gamma_-+\phi/2)}\\
\end{pmatrix}.
\end{align}
Note that in writing down these matrices, we have explicitly separated out the superconducting U(1) phase corresponding to the broken symmetry and defined $-\i\alpha_\pm = \e{\i\delta_\pm}$. Without loss of generality, we set it to $\pm\phi/2$ on the right and left side, respectively. The phase-factors $\gamma_\pm$ thus only contain information about the internal phase in $\vk$-space for the order parameter.

\end{widetext}

\end{document}